\documentclass[12pt]{article}
\usepackage{amssymb}
\usepackage{amsthm}
\usepackage{amsmath} 
\usepackage{xcolor}

\usepackage{tikz}

\usetikzlibrary{arrows}
\usepackage{stmaryrd,bm}
\usepackage{accents}
\usepackage[normalem]{ulem} 
\usepackage{mathrsfs}
\usepackage{enumitem}
\usepackage{isomath}
\usepackage{rotating}
\usepackage{cite}
\usepackage[unicode,linkcolor=blue,colorlinks=true,citecolor=red]{hyperref}
\usepackage{soul}

\newcommand{\bb}{\bm b}
\newcommand{\dst}{\displaystyle}
\renewcommand{\leq}{\leqslant}
\renewcommand{\geq}{\geqslant}
\newcommand{\noin}{\noindent}

\renewcommand{\tilde}{\widetilde}
\newcommand{\ep}{\varepsilon}
\renewcommand{\phi}{\varphi}
\newcommand{\kp}{k}

\newcommand{\mb}{\bm m}

\newcommand{\br}{\bm x}

\newcommand{\bk}{{\bm k}}

\newcommand{\n}{{\bm n}}
\newcommand{\de}{\partial}
\newcommand{\x}{{\bm x}}
\newcommand{\bxi}{{\bm \xi}}
\newcommand{\et}{{\bm \eta}}
\newcommand{\y}{{\bm y}}

\newcommand{\hpsi}{\psi}
\newcommand{\rf}[1]{(\ref{#1})}
\newcommand{\ut}{\tilde{u}}

\newcommand{\walpha}{\what{\alpha}}

\renewcommand{\setminus}{\smallsetminus}

\newcommand{\what}{\widehat}
\newcommand{\nonu}{\nonumber}
\newcommand{\oh}{\frac{1}{2}}
\renewcommand{\th}{\frac{3}{2}}

\newcommand{\z}{{\bm 0}}

\newcommand{\A}{\mathsfbfit A}
\newcommand{\Am}{\bm A}
\newcommand{\B}{\mathsfbfit B}
\newcommand{\Csbi}{\Cm}
\newcommand{\C}{\mathscr C} 
\newcommand{\Cm}{\bm C}
\newcommand{\Dsbi}{\mathsfbfit D}
\newcommand{\Eb}{\bm E}
\newcommand{\Dm}{\bm D}
\newcommand{\Es}{{\mathsfbfit E}}
\newcommand{\F}{{\mathsfbfit F}}
\renewcommand{\H}{{\mathsfbfit H}}
\newcommand{\K}{{\mathscr K}}
\newcommand{\T}{{\mathsfbfit T}}
\newcommand{\Th}{\what{\T}}
\newcommand{\Hh}{\what{\H}}
\newcommand{\Ho}{H^\oh}
\newcommand{\Ht}{H^\frac{3}{2}}

\newcommand{\M}{\bm M}
\newcommand{\Msbi}{\M}

\newcommand{\Om}{\Omega}
\newcommand{\Omh}{\what{\Omega}}

\newcommand{\R}{{\mathbb R}}
\newcommand{\U}{{\mathsfbfit S}} 

\newcommand{\Id}{\mathsfbfit I}
\newcommand{\Ih}{\what{\mathsfbfit I}}
\renewcommand{\Im}{\bm I}
\newcommand{\J}{\bm J}

\renewcommand{\O}{{\cal O}}

\newcommand{\N}{\mathsfbfit N}
\newcommand{\Np}{{\N}^{+}}
\newcommand{\Nm}{{\N}^{-}}

\newcommand{\om}{\omega}

\newcommand{\ei}{\varepsilon}

\newtheorem{theorem}{Theorem}[section]

\newtheorem{lemma}{Lemma}[section]
\newtheorem*{definition}{Definition}

\newtheorem{remark}{Remark}

\addtocounter{figure}{0}

\newcommand{\I}{\ensuremath{\mathrm{i}}}
\newcommand{\E}{\ensuremath{\mathrm{e}}}
\newcommand{\D}{\ensuremath{\mathrm{d}}}

\begin{document}

\title{Propagation and dispersion of Bloch waves \\ in periodic media with soft inclusions}
\author{
Yuri~A. Godin\thanks{Email: ygodin@uncc.edu} and Boris Vainberg\thanks{Email: brvainbe@uncc.edu} \\
The University of North Carolina at Charlotte, \\
Charlotte, NC 28223 USA
}

\maketitle

\begin{abstract}
We investigate the behavior of waves in a periodic medium containing small soft inclusions or cavities of arbitrary shape, such that the homogeneous Dirichlet conditions are satisfied at the boundary. 
The leading terms of Bloch waves, their dispersion relations,  and cutoff frequencies are rigorously derived.
Our approach reveals the existence of exceptional wave vectors for which Bloch waves are comprised of clusters of perturbed plane waves that propagate in different directions. 
  We demonstrate that 
  for these exceptional wave vectors, 
  no Bloch waves propagate in any one specific direction. 
\end{abstract}



\bigskip

\section{Introduction}
\setcounter{equation}{0}

The study of wave propagation in periodic media has led to numerous practical discoveries, 
including the appearance of bands and gaps in the wave spectrum, 
nonreciprocal media, negative group velocity, 
and the self-collimation effect. 
These phenomena have resulted in the development of new devices.  Various numerical methods, such as plane wave expansion, finite-difference, finite-element, and boundary-element methods, have been used to study the dispersion of waves in periodic structures, see references in \cite{JJWM:11, Kuchment:99}. Although deriving an explicit dispersion relation for the Floquet-Bloch waves in two and three-dimensional periodic media is a difficult problem, an asymptotic approximation can be obtained by assuming that the wavelength is long compared to the period of the lattice or the characteristic size of the scatterers \cite{Kuchment:99a}. One commonly used technique to study dispersion relations in periodic media is matched asymptotic expansions, which have been applied to small inclusions with Dirichlet or Neumann boundary conditions \cite{McIver:07, McIver:2009, McIver:2011, Guo:14, Craster:2017,  Craster:20}.

In \cite{MMP:02}, a semi-analytical method based on the multipole expansion technique is discussed. In \cite{GV:19}, formulas for the dispersion relation and effective dielectric tensor are derived under the assumption that the cell size is small relative to the wavelength but large relative to the size of the inclusions. Other approaches can be found in \cite{Zalipaev:02,Craster:10,Vanel:17,Cherednichenko:06,Smyshlyaev:08,Joyce:17,Guzina:21}. However, most previous works rely on the construction of waves that approximately satisfy the equation and boundary conditions. Without an estimate on the inverse operator, it cannot be ensured that the constructed functions are close to the exact solution.

We considered the transmission problem in  \cite{GV:22}. It concerns  
the propagation of acoustic waves in an infinite medium containing a periodic array of identical inclusions of arbitrary shape. The ``hard'' inclusions with the Neumann boundary condition is a particular case of the situation studied in \cite{GV:22} when the mass density $\varrho$ of the inclusion $\varrho \to \infty$  while the adiabatic bulk compressibility modulus $\gamma \to 0$. 
We obtained not just a function that almost satisfies the equation, but an asymptotic expansion of the exact solution. The rigorous analysis allowed us to justify the presence of exceptional wave vectors, for which every solution to the problem takes the form of a cluster of waves propagating in different directions, with no one of them propagating in a single direction. This phenomenon is somewhat similar to the Bragg reflection \cite{Brillouin:56} and wave refraction. It appears in solid-state physics in the study of energy gaps of electrons in the field of a weak periodic potential \cite{AM:76, Kittel:95, Karpeshina:97}.

The main focus of this paper is on acoustic waves in an infinite medium that has a periodic array of soft inclusions or cavities of arbitrary shape with the Dirichlet boundary condition. This problem is distinct from the one studied in \cite{GV:22} as it cannot be obtained as a particular case of the transmission problem. To tackle this new problem, we have modified our previous approach. The problem with cavities is more challenging than the transmission problem, and we have only been able to develop an asymptotic expansion of dispersion relations and solutions for non-exceptional wave vectors and for some clusters when the wave vector is exceptional.  
Unlike the transmission problem, the dispersion relations of clusters in the case of cavities have asymptotics of different orders. 


The difficulty related to small inclusions is because the latter problem is a singular perturbation of the problem without inclusions. Instead of tackling the daunting task of studying this singularly perturbed problem, we simplify it by transforming it into an operator equation on the surface of a fixed sphere enclosing the cavity. This transformation enables us to apply standard perturbation theory and determine the solution to the auxiliary problem in the form of a power series, which we can then use to rigorously construct the solution to the original problem.

The paper is structured as follows: \S \ref{form} outlines the problem and introduces the concept of the exceptional Bloch vector, as well as provides a brief overview of the main results. \S \ref{out} constructs the interior and exterior Dirichlet-to-Neumann (DtN) operators on the sphere of a fixed radius $R$ and transforms the original problem into a study of the zero eigenvalue of the difference of the DtN operators where regular perturbation theory can be applied. \S\S \ref{exter} and \ref{inner} present the expansion of the DtN operators in power series, using small parameters of the problem. In \S \ref{main}, the matrix of the interior operator is determined in the basis of unperturbed Bloch functions. The main theorems on the solution structure and dispersion relations are formulated and proven in \S \ref{pt1}. Finally, a conclusion and a discussion of the further line of research are given in \S \ref{concl}.

\section{Formulation of the problem and main results}
\setcounter{equation}{0}
\label{form}

We study the behavior of time-harmonic acoustic waves propagating through a three-dimensional periodic lattice consisting of identical cavities, denoted as $\Om$. The amplitude $u$ of these waves  is described by an equation
\begin{align}
 \Delta u + k^2 u =0
 \label{eq1}
\end{align}
and the Dirichlet boundary condition $u=0$ on the boundary of the cavities.
Here $k=\om/c$ is the wave number, $\om$ is time frequency and $c$ is the speed of wave propagation in the host medium.

We are looking for a solution $u$  of the problem in the form of Bloch waves
\begin{align}
 u(\x) = \Phi(\x) \E^{-\I \bk \cdot \x},
 \label{fbw}
\end{align}
where $\bk = (k_1, k_2, k_3)$ is the wave vector, $k = |\bk|$, and function $\Phi$ is periodic with the periods of the lattice. The latter can be rewritten as
\begin{equation}
 \rrbracket  \E^{\I\bk \cdot \x} u(\x) \llbracket =0,
\label{fb1}
\end{equation}
where $\rrbracket f \llbracket$ denotes the jump of $f$ and its gradient across the opposite sides of the cells of periodicity.

Thus, the problem in the medium is reduced to the following one in the fundamental cell of periodicity $\Pi$:
\begin{align}
\label{Hz1}
 \Delta u + k^2 u &=0, \quad u \in H^2 (\Pi \setminus \Om), \\[2mm]
  \left. u \right|_{\de\Om} &= 0, \quad  \rrbracket  \E^{\I\bk \cdot \x} u(\x) \llbracket =0.
 \label{bc1a}
\end{align}
Here $H^2$ is the Sobolev space and $\Om$ is the domain occupied by the cavity in $\Pi$.

\begin{figure}[th]
\begin{center}
\begin{tikzpicture}[scale=1.1,>=triangle 45]
 \begin{scope}[x={(4cm,0cm)},y={({cos(30)*2.5cm},{sin(30)*2.5cm})},
    z={({cos(70)*3cm},{sin(70)*3cm})},line join=round,fill opacity=0.5]
  \draw[fill=lightgray!50] (0,0,0) -- (0,0,1) -- (0,1,1) -- (0,1,0) -- cycle;
  \draw[fill=lightgray!50] (0,0,0) -- (1,0,0) -- (1,1,0) -- (0,1,0) -- cycle;
  \draw[fill=lightgray!50] (0,1,0) -- (1,1,0) -- (1,1,1) -- (0,1,1) -- cycle;
  \draw[fill=lightgray!50] (1,0,0) -- (1,0,1) -- (1,1,1) -- (1,1,0) -- cycle;
  \draw[fill=lightgray!50] (0,0,1) -- (1,0,1) -- (1,1,1) -- (0,1,1) -- cycle;
  \draw[fill=lightgray!50] (0,0,0) -- (1,0,0) -- (1,0,1) -- (0,0,1) -- cycle;
  \draw [->,very thick] (0,1,0) -- (1,1,0);
  \draw [->,very thick] (0,1,0) -- (0,0,0);
  \draw [->,very thick] (0,1,0) -- (0,1,1);
 \end{scope}
 \draw [->, thick] (4.15,1.55) -- (5.05,1.55) node [above right] {\Large  $x_2$};
 \draw [->, thick] (2.85,1.05) -- (2.35,0.35) node [left] {\Large  $x_1$};
 \draw [->, thick] (3.35,2.55) -- (3.35,3.35) node [above right] {\Large  $x_3$};
 \shade[ball color = teal!10!white, opacity = 0.9] (4.5,2.9,3) circle (1cm);

 \def\eggheight{4mm}
 \begin{turn}{45}
  \path[ball color=white, opacity = 1]
  plot[domain=-pi:pi,samples=100,shift={(3.58,-1.15)}]
  ({.8*\eggheight *cos(\x/4 r)*sin(\x r)},{-\eggheight*(cos(\x r))})
  -- cycle;
  \end{turn}
  \node [above] at (3.35,1.45) {$\Om$};
  \node [above] at (5.8,4.0,2) {\Large$\Pi$};
  \node [above] at (3.2,1.99) {\large$B_R$};
  \node [above] at (6.4,0.5) {\Large $\bm \ell_2$};
  \node [above] at (-0.3,0) {\Large $\bm \ell_1$};
  \node [above] at (3,4.1) {\Large $\bm \ell_3$};
\end{tikzpicture}
\end{center}
\caption{
The cell of periodicity $\Pi$ with a small cavity $\Om$ inside. A ball $B_R \subset \Pi$ of radius $R$ centered at the origin encloses the cavity $\Om$.
}
\label{cell}
\end{figure}
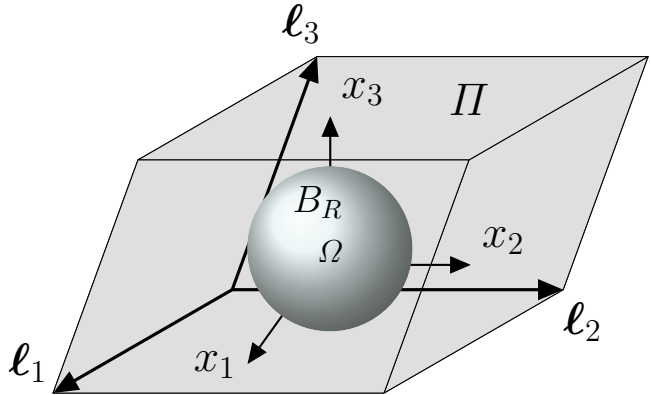
Let us select the origin for the coordinate system in $\Om$ and suppose that $\Om$ has a small size. Specifically, $\Om = \Om (a)$ is produced by compressing an $a$-independent region $\what{\Om}$ using the factor $a^{-1}$, $0 < a \ll 1$. In other words, the transformation $\x \to a\bxi $ maps  $\Om (a) \subset \R^3_{\x}$ into $\Omh \subset \R^3_{\bxi}$. We assume that the boundary $ \de\Om$ belongs to the class  $C^{1,\beta}$, which means that the functions that define the boundary have first-order derivatives that belong to the H\"{o}lder space with index $\beta$.

We fix the shape $\Omh$ of the cavities and consider the dependence of the solution on all other parameters including the size $a$ of the cavity.
A non-trivial solution of  \rf{Hz1},\rf{bc1a} with a fixed $\Omh$ does not exist for all the values of $k=\om/c, a,$ and $\bk$. The relation between the parameters $\om, a, \bk$ for which such a solution exists is called a dispersion relation.

Using the conventional method, we introduce the basis vectors $\bb_1, \bb_2, \bb_3$ of the reciprocal lattice $\mathbb Z^3_b$
\begin{align}
 \label{}
 {\bm \ell}_i \cdot \bb_j = 2\pi \delta_{i,j},
\end{align}
where $\delta_{i,j}$ is the Kronecker delta. Points $\mb$ on the lattice can be expressed as 
\begin{align}
 \mb = m_1 \bb_1 + m_2 \bb_2 + m_3 \bb_3,
 \label{mb}
\end{align}
where $m_1, m_2, m_3$ are integers. If $\Pi$ is a cube $[-\pi,\pi]^3$, then $\mathbb Z^3_b$ refers to the standard lattice $\mathbb Z^3$ of points $\mb = (m_1 , m_2,  m_3 )$. It is important to note that the space of Bloch waves is not always one-dimensional, even for the unperturbed problem.

\begin{definition}
 A point $\bk$ is called exceptional of order $n$ if there are non-trivial vectors $\mb = \mb_s, ~ 2 \leq s \leq n,$ of the form \rf{mb} such that $|\bk| = |\bk - \mb|$. We also set $\mb_1 = \z.$ 
\end{definition}

\begin{remark}
 Geometrically, $n$ refers to the number of points in the reciprocal lattice $\mathbb Z^3_b$ (including the origin) that belong to the sphere centered at $\bk$ with a radius of $|\bk|$. If the origin (which is always on the sphere) is the only point of the reciprocal lattice on the sphere, the point $\bk$ is non-exceptional. This is why the order of exceptional points starts with $n=2$.
\end{remark}

Consider the unperturbed problem
\begin{equation}
\label{unp}
\Delta u + |\bk|^2  u = 0, \quad u \in H^2 (\Pi), \quad
\rrbracket  \E^{\I\bk \cdot \br} u(\br) \llbracket =0.
\end{equation}
 \begin{lemma}\label{lex}
 If $\bk$ is a non-exceptional wave vector, then the solution space of the unperturbed problem \rf{unp} consists of functions $C \E^{-\I \bk \cdot \x}$. However, if $\bk$ is an exceptional wave vector of order $n$, then the solution space is $n$-dimensional and can be expressed as a linear combination of functions $\psi_s=\E^{-\I (\bk -\mb_s)\cdot \x}, ~ 1\leq s\leq n.$ Here, $\mb_s $ are points on the reciprocal lattice $\mathbb Z^3_b$ that satisfy $|\bk| = |\bk - \mb_s|$.
 \end{lemma}
 \begin{proof}
We first remind that the functions $\psi_s, ~ 1\leq s\leq n,$ satisfy \rf{unp}, and it remains to show that there are no other linearly independent solutions. 
Suppose that $u$ satisfies \rf{unp}. Then we construct a periodic solution $v(\x)=\E^{\I \bk \cdot \x}u(\x)$ of the equation $\Delta v -2i\bk\cdot \nabla v = 0$. This function $v$ can be represented as a Fourier series
\[
v(\x)=\sum_{\mb\in\mathbb Z_b^3}a_{\mb}\E^{-\I \mb\cdot \x},
\]
and upon substitution into the equation, we obtain the relation $(2\bk \cdot \mb-|\mb|^2)a_{\mb}=0$. This shows that $a_{\mb}$ can only be non-zero if $2\bk \cdot \mb-|\mb|^2=0$, which is equivalent to $|\bk| = |\bk - \mb|$. Hence, $v(\x)=a_{\z}$ is a constant if $\bk$ is not exceptional, and $v(\x)$ is a linear combination of functions $\E^{\I \mb_s\cdot \x},~1\leq s\leq n,$ if $\bk$ is an exceptional vector. Thus, $u$ is a linear combination of $\psi_s$.
\end{proof}

 The order of an exceptional wave vector can be understood in terms of the planes defined by the equation $2\bk \cdot \mb=|\mb|^2$, where $\mb \neq \z$ belongs to the reciprocal lattice $\mathbb Z^3_b$, and $\bk$ is a wave vector. If $\bk$ belongs to at least one of these planes, it is considered exceptional. The order of an exceptional wave vector is equal to the number of the planes to which $\bk$ belongs plus one. 
 As the magnitude of $\mb$ increases, so does the distance from the origin to these planes.
 
 It has been known that the space of Bloch waves in the unperturbed problem can be multidimensional (when $\bk$ is an exceptional point). However, this result is underestimated because the basis in this multidimensional space can be chosen in such a way that each Bloch wave propagates in its direction. The latter created a general belief that the same is true for the perturbed problem. One of the important results of \cite{GV:22} and the present paper is that for the problem with small inclusions or cavities, each solution satisfying the Bloch condition is a cluster of plane waves propagating in different directions with different spatial frequencies. Moreover, there are no Bloch solutions propagating in one specific direction.  
 
Upon the introduction of a small perturbation, the Bloch wave featuring a non-exceptional $\bk$ undergoes a slight change. We will demonstrate that the relation between $\om$ and $\bk$ varies smoothly concerning $a$, thereby establishing a well-defined dispersion relation. The asymptotic behavior of the dispersion relation has the form

\begin{align}
\frac{\om^2}{c^2} = |\bk|^2 + \frac{4\pi a q}{|\Pi|} + \O(a^2), \quad a\to 0,
\label{om_reg}
\end{align}
where $c$ is the speed of waves,  $|\Pi|$ denotes the volume of the cell of periodicity $\Pi$, and the coefficient $q$ depends on the shape of the cavity and is equal to the total charge of the surface $\de \Omh$ whose potential is one, i.e. $aq$ is the capacitance of $\Om$ ($q=1$ for a sphere). This Bloch wave propagates in the direction of vector $\bk$.

When $\bk$ is an exceptional wave vector of order $n$, the presence of cavities produces several {\it clusters} consisting of waves propagating in the directions of $\bk-\mb_j, 1\leq j\leq n$. 
One of them has the form (details are given in Theorem \ref{t4})
 \begin{align}
 u_1(\x) = C_1 \sum_{j=1}^n \E^{-\I (\bk - \mb_j) \cdot \x} + \O(a),
 \quad a \to 0,
\end{align}
with the dispersion relation
\begin{align}
  \left(\frac{\om_1}{c}\right)^2 = |\bk|^2 + \frac{4\pi a q n}{|\Pi|} + \O(a^2), \quad a\to 0.
 \label{om_excep}
\end{align}
If $n=2$ then 
\begin{align}
 u_2(\x) = C_1 \left( \E^{-\I (\bk - \mb_1) \cdot \x}  - \E^{-\I (\bk - \mb_2) \cdot \x} \right)+ \O(a), \quad\
 \left(\frac{\om_2}{c}\right)^2 = |\bk|^2 + \O(a^2), \quad a\to 0.
\end{align}
When $n >2$ then the term of order $\O(a)$ vanishes in the dispersion relations for all clusters except $u_1$.

{\it
No cluster $u_s$ propagates in one specific direction $\bk-\mb_j,~1 \leq j\leq n$.
A linear combination of clusters $u_s$ may propagate in one direction, but such a combination does not have mathematical or physical meaning since the terms would satisfy equations \rf{Hz1} with different values of $k=k_s$ and the time frequencies $\om_s = k_s/c, ~s\geq 1$ of these terms are distinct. 
}

Finally, the solution of \rf{Hz1} and\rf{bc1a} does not exist if $\om$ is less than a cutoff frequency $\om_c$. We show that for non-exceptional wave vectors and clusters with dispersion relation \rf{om_excep}, the cutoff frequency $\om_c$ is given by
\begin{align}
 \om_c = 2c\, \sqrt{\frac{\pi a n q}{|\Pi|}} \left( 1 + \O(a)\right), \quad a \to 0,
\end{align}
and $\om_c = \O(a)$ for other clusters.

\section{Outline of the approach}
\label{out}
\setcounter{equation}{0}

In this paper, we adjust our approach introduced in \cite{GV:22} to the cavities with the Dirichlet boundary conditions. 
In a homogeneous medium, the dispersion relation has the form $k = \om/c = |\bk|$ with nontrivial solutions of the problem in the form of plane waves  $u = \E^{\I \bk \cdot \x}$.
For small cavities, we will be looking for the dispersion relation in the form
$ k=\om/c = (1+\ep) |\bk|$, where $\ep = \ep (a, |\bk|)$ vanishes as $a \to 0$. We need to determine the asymptotic behavior of $\ep$ as $a \to 0$.

The problem described by equations \rf{Hz1},\rf{bc1a} presents a challenge as it pertains to a singular perturbation of the inclusionless medium, which results in an intricate behavior of the solution near the cavity. To derive the dispersion relation, we use an auxiliary regularly perturbed problem. This involves enclosing the cavity $\Om = \Om(a)$ within a ball $B_R \subset \Pi$ of radius $R > a$,  and centered at the origin. We then divide the region $\Pi$ into two separate components, namely the ball $B_R = \{ |{\x} | < R\}$ and its complementary $\Pi \setminus B_R$ (refer to Figure \ref{cell}), and address two distinct problems within each of these regions.
 \begin{align}
 \left( \Delta + k^2 \right) u(\x) &= 0, \quad \x\in B_R \setminus \Om, \quad
 \left. u \right|_{\de \Om}=0, \quad\left. u\right|_{r=R}=\psi.
 \label{uin}\\[2mm]
 \label{k-egv}
 \left( \Delta + k^2 \right) v(\x) &= 0, \quad  \x\in \Pi \setminus B_R,~~  \left\rrbracket  \E^{\I \bk \cdot \x} v(\x) \right\llbracket =0, ~~ \left. v\right|_{r=R}=\psi, \quad k = (1+\varepsilon) |\bk|.
\end{align}

In the unperturbed case, $\varepsilon =0, ~ a=0$, problems \rf{uin}, \rf{k-egv} are uniquely solvable for all values of $R$, except possibly a discrete set $\{R_i\}$. We fix an $R\notin \{R_i\}$. It will be shown that solutions of \rf{uin}, \rf{k-egv} are still unique for fixed $R\notin \{R_i\}$ when $a$ and $\varepsilon$ are small. We introduce Dirichlet-to-Neumann (DtN) operators $\Nm_{a,\varepsilon}$, $\Np_{\bk,\varepsilon}$ for problems \rf{uin}, \rf{k-egv}, respectively,  with the derivatives in the direction of $r$:
\begin{equation}
\label{}
\Nm_{a,\varepsilon}, \Np_{\bk,\varepsilon}:\Ht(\de B_R)\to \Ho(\de B_R),
\quad~~ \Nm_{a,\varepsilon}  \psi = \left.\frac{\de u}{\de r}\right|_{r=R}, \quad  \quad
\Np_{\bk,\varepsilon}  \psi = \left.\frac{\de v}{\de r}\right|_{r=R},
\end{equation}
where $H^{s}(\de B_R)$ is the  Sobolev space of functions on $\de B_R$.

The method developed in this paper exploits essentially the following two lemmas.

\begin{lemma}\label{fred}
 Operators
 \begin{equation}\label{npm}
\Np_{\bk,\varepsilon},~\Nm_{a,\varepsilon},~~ \text{and} ~~ \Np_{\bk,\ep}-\Nm_{a,\ep}:\Ht(\de B_R)\to \Ho(\de B_R)
 \end{equation}
are Fredholm and symmetric in $L^{2}(\de B_R)$. Thus, for example,
 \begin{equation}\label{sym}
  \int_{r=R} \left( \Np_{\bk,\ep} \psi \right) \overline{\phi}\, \D S = \int_{r=R} \psi \, \overline{\left(\Np_{\bk,\ep} \phi\right)}\, \D S,  \quad \psi,\phi\in \Ht(\de B_R).
 \end{equation}

\end{lemma}
The proof can be found in \cite{GV:20} where the precise shape of $\Om$ was not used. The Fredholm character of the operators is a result of their ellipticity (the symbols of the DtN operators and their ellipticity can be found in \cite{vg}). The symmetry easily follows from Green’s formula.

\begin{lemma}\label{one-to-one}
\leavevmode
Relation $\psi=u|_{r=R}$ is a one-to-one correspondence between solutions $u$ of \rf{Hz1}-\rf{bc1a} and solutions $\psi\in \Ht(\de B_R)$ of
\begin{align}
\label{nn}
(\Np_{\bk,\varepsilon}-\Nm_{a,\varepsilon})\,\psi=\z.
\end{align}
\end{lemma}
\begin{proof}
Suppose $u$ satisfies equations \rf{Hz1} and \rf{bc1a}. In that case, the Dirichlet and Neumann data for the solution of \rf{uin} coincide with those for the solution of \rf{k-egv}, and we can conclude that \rf{nn} is valid. Conversely, if we extend any solution of \rf{uin} with $\psi$ satisfying \rf{nn} onto the domain $\Pi \setminus B_R$ using the solution of \rf{k-egv}, we obtain a solution of \rf{Hz1} and \rf{bc1a}, provided that the Dirichlet and Neumann data on $\de B_R$ of the solution of \rf{uin} match with those of the solution of \rf{k-egv}.
\end{proof}
Lemma \ref{one-to-one} plays a crucial role in our paper, as it reduces the dispersion relation to the dependence between $\om$ and $\bk$ which results in the existence of a zero eigenvalue of $\Np_{\bk,\varepsilon} - \Nm_{a,\varepsilon}$. Our approach has a significant advantage since it allows us to determine the dispersion relation using standard perturbation theory, given that $\Np_{\bk,\varepsilon}$ is independent of the inclusion, and $\Nm_{a,\varepsilon}$ is an infinitely smooth operator function of $a$.  For convenience, we will replace $\Np_{\bk,\varepsilon}$ and $\Nm_{a,\varepsilon}$ with $\Np_{\bk,\varepsilon}- \Nm_{0,\varepsilon}$ and $\Nm_{a,\varepsilon} - \Nm_{0,\varepsilon}$, respectively, as we proceed.

\section{Expansion of the exterior DtN operator}
\label{exter}
\setcounter{equation}{0}

\begin{lemma}
\label{l3a}
In the absence of cavities, the operator $(\Np_{\bk,0}-\Nm_{0,0}): \Ht(\de B_R) \to\Ho (\de B_R)$ has a simple zero eigenvalue for a non-exceptional vector $\bk$, while for an exceptional vector $\bk$, the zero eigenvalue has a multiplicity $n$, where $n$ is the order of the exceptional vector. The space of eigenfunctions is spanned by the functions $\psi_{s}$ defined by
\begin{equation}\label{psi0}
\psi_{s}:=\E^{-\I (\bk - \mb_s) \cdot \x}|_{r=R}, \quad 1 \leq s \leq n,
\end{equation}
where $\mb_1 = \z$ and the points $\mb_s\in \mathbb Z^3_b$, $s > 1$, were introduced in the definition of the exceptional point. 
\end{lemma}
\begin{proof}
Lemma \ref{lex} shows that the solution space of the inclusionless problem \rf{Hz1},\rf{bc1a} consists of linear combinations of functions \rf{psi0}. Lemma \ref{one-to-one} implies that this solution space and the kernel of the operator $(\Np_{\bk,0}-\Nm_{0,0})$ coincide.
\end{proof}

We denote the finite-dimensional space spanned by functions $\psi_{s}$, $1\leq s \leq n$ as $\mathscr E$, and we define $\mathscr E_{\bot,1}$ and $ \mathscr E_{\bot,0}$ as the subspaces of $\Ht(\de B_R)$ and $\Ho (\de B_R)$, respectively, consisting of functions orthogonal in $L^2(\de B_R)$ to $\mathscr E$. We represent each element $\psi$ in the domain $\Ht(\de B_R)$ and the range $\Ho (\de B_R)$ of operator \rf{npm} in the vector form $\psi=( \psi_{\mathscr E},\psi_\bot),$ where $\psi_{\mathscr E}$ is the projection in $L^2(\de B_R)$ of the function $\psi$ into the space $\mathscr E$, and $\psi_\bot$ is orthogonal to $\psi_{\mathscr E}$ in $L^2(\de B_R)$. Using Lemmas \ref{fred} and \ref{l3a}, the unperturbed operator \rf{npm} can be written in matrix form:

\begin{equation}
\label{A}
 \Np_{\bk,0} - \Nm_{0,0} = \left(
 \begin{array}{cc}
  0 & 0 \\[2mm]
  0 & \Am
 \end{array}
 \right),
\end{equation}
where $\Am:\mathscr E_{\bot,1}\to \mathscr E_{\bot,0}$ is an isomorphism.

As equation \rf{k-egv} is a smooth function of both $\bk$ and $\ep$, where $|\ep|<1$, the operator $\Np_{\bk,\varepsilon}$ is infinitely smooth in both its arguments. It should be noted that $|\ep|$ becomes small as $a$ becomes small, as discussed in \S \ref{out}. Therefore, in the basis $( \psi_{\mathscr E},\psi_\bot)$, the matrix representation of the operator $\Np_{\bk,\ep} - \Nm_{0,\ep}$ takes the following form as $\ep$ approaches zero:
\begin{equation}
\label{conc}
 \Np_{\bk,\varepsilon} - \Nm_{0,\varepsilon} = \left(
 \begin{array}{cc}
  \Cm\ei + O(\ei^2) & O(\ei) \\[2mm]
  O(\ei) &\Am + O(\ei)
 \end{array}
 \right)=\left(
 \begin{array}{cc}
  \Cm\ei + \ei^2 \Dm_{11} & \ei \Dm_{12} \\[2mm]
  \ei \Dm_{21} & \Am + \ei \Dm_{22}
 \end{array}
 \right), \quad \ei=(\kp -|\bk|)/|\bk|,
\end{equation}
where $\Cm$: $\mathscr E \to \mathscr E$ is a finite-dimensional operator,  operators $\Dm_{ij}=\Dm_{ij}(\ei,\bk)$ are infinitely smooth functions of the arguments, and equation for $\ep$ is equivalent to that in \rf{k-egv}. We fix the basis \rf{psi0} in $\mathscr E$ and identify operator $\Cm$ with its matrix representation in this basis. Matrix $\Cm$ does not depend on the interior problem and has been determined in \cite{GV:22}. Its evaluation is based on Green's formula and the result is as follows.
\begin{lemma}
\label{l33}
Operator $\Cm$ in \rf{conc} has the form
 \begin{equation}
 \label{C}
  \Cm = 2|\bk|^2 |\Pi| \Im,
 \end{equation}
 where $\Im$ is the identity operator and $|\Pi|$ is the volume of $\Pi$.
\end{lemma}

The next two sections are devoted to the asymptotics of the interior DtN operator. First, it is done for an arbitrary boundary condition on $\de B_R$ and then in a special basis related to the formulas \rf{A},\rf{conc}.

\section{Asymptotics of the interior DtN operator}
\setcounter{equation}{0}
\label{inner}

Consider the Dirichlet problem in $B_R \setminus \Om$:
\begin{align}
\label{5.1}
 \Delta u + k^2 u &=0, \quad u \in H^2 (B_R \setminus \Om), \\[2mm]
  \left. u \right|_{\de\Om} &= 0, \quad  \left. u \right|_{\de B_R} = \psi \in H^\th (B_R \setminus \Om).
 \label{5.2}
\end{align}
The asymptotic expansion of the solution $u$ for small values of $a$, in a neighborhood of the origin, is quite complicated. We will not delve into this asymptotic behavior in this study, though it can be obtained from the reasoning presented below.
However, a simple power series representation for the solution $u$ of \rf{5.1}, \rf{5.2} exists when $|\x| > \delta >0$ and for the DtN map: $\Nm_{a,\varepsilon}: \left.\psi=u \right|_{\x\in B_R}\to \left.\frac{\de u }{\de \n}\right|_{\x\in B_R}$ and they can be obtained without studying the asymptotics of $u$ around the origin. 

The main result of this section is given by Theorem \ref{t1}. We introduce some notation to formulate the theorem.

Solution $u$ can be presented in the form
\begin{align}
 u = u_0 + \ut,
 \label{}
\end{align}
where $u_0$ is the solution of the problem in the ball $B_R$ without a cavity:
\begin{align}
\Delta u_0 + k^2 u_0 = 0, \quad \x \in B_R,  \quad  \left. u_0 \right|_{\de B_R} = \psi.
 \label{u0}
\end{align}
Then $\ut$ satisfies
\begin{align}
\label{eq_ut}
\Delta \ut + k^2 \ut &= 0, \quad \ut \in H^2 (B_R \setminus \Om), \\[2mm]
 \left. \ut \right|_{\de\Om} &= -u_0, \quad  \left. \ut \right|_{\de B_R} = 0.
 \label{bc_ut}
\end{align}

For the problem without a cavity, the image of the DtN map is equal to $\Nm_{0,\varepsilon} \psi = \left.\frac{\de u_0 }{\de \n}\right|_{\x\in B_R}$, and our goal is to find the asymptotics of $\ut$ and of the difference
\begin{equation}\label{dn}
\Nm_{a,\varepsilon}-\Nm_{0,\varepsilon}:\psi\to \left.\frac{\de \ut }{\de \n}\right|_{\x\in B_R}.
\end{equation}

For the formulation of the main result of this section, we introduce the notation $q$ for the total electric charge 
of the conducting surface $\de \Omh$ for which the electric potential $v$ on $\de \Omh$ equal to one.
It can be found from the solution to the problem
\begin{align}
\Delta v_\bxi = \frac{\de^2 v}{\de \xi_1^2}  + \frac{\de^2 v}{\de \xi_2^2} + \frac{\de^2 v}{\de \xi_3^2} = 0,
 \quad \bxi \in \R^3 \setminus \Omh, \quad
\left. v \right|_{\de \Omh} = 1, \quad v=\O(|\bxi|^{-1})  \quad
{\rm as} ~~
\bxi\to\infty.
 \label{v}
\end{align}
It is well known that $v=q/|\bxi|+\O(|\bxi|^{-2}),~\bxi\to\infty,$ where $q$ is a constant.
From Green's formula, it follows that $q = -\frac{1}{4\pi}\int_{\de \Omh} \frac{\de v}{\de \n}\, \D S$, i.e. the coefficient $q$ in the asymptotics of $v$ at infinity coincides with the total charge of the surface $\de \Omh$ with unit potential. 

\begin{theorem}
\label{t1}
Let $R \notin \{R_i\}$. Then for arbitrary $\delta>0$ there exist functions $\ut_j = \ut_j(\ep,\x) \in C^\infty$ such that 
\begin{align}
\ut=\sum_{j=1}^\infty \ut_j(\ep,\x)\,a^j,  \quad a\to 0, \quad \ut_1 &= -\frac{q u_0 (\z)}{ |\x|}\left(\cos k|\x| - \cot kR\, \sin k|\x| \right),
\label{uwa}
\end{align}
where  $\delta\leq|\x|\leq R,~|\ep|\ll 1,$ and the series converges in $C^N$ for any $N>0$.  Therefore
\begin{align}
\left(\Nm_{a,\varepsilon}-\Nm_{0,\varepsilon}\right) \psi =  \frac{\de \ut}{\de \n} &=\sum_{j=1}^\infty \frac{\de \ut_j(\ep,\x)}{\de \n} \,a^j=\frac{a q k u_0 (\z)}{R\sin kR} + \O (a^2), \quad a \to 0, \quad \x\in B_R .
 \label{Nam}
\end{align}
\end{theorem}
\begin{remark}
 The smoothness in $\ep$ follows from the smooth dependence of the solution of \rf{Hz1}, \rf{bc1a} in $k$, and by default, we omit the dependence of the solutions on $\ep$.
\end{remark}

The proof of the theorem is given at the end of this section and it is based on the introduction of a double layer potential associated with the Helmholtz equation. We look for the solution of \rf{eq_ut},\rf{bc_ut} in the form 
\begin{align}
 \ut  = \H \alpha, \quad \alpha \in H^\oh (\de \Om),
 \label{ut}
\end{align}
where $\H$ is the double layer potential 
\begin{align}
\label{H}
(\H \alpha ) (\x) &= \int_{\de \Om}  H(\x,\y)  \alpha (\y)\,\D S_\y, \quad \H: H^\oh (\de \Om) \to H^2 (B_R \setminus  \Om), \quad \y \in \de \Om.
\end{align}
with the kernel that is a modification of the standard one:
\begin{align}
 H(\x,\y) = \frac{\de E(\x,\y)}{\de \n_\y}-F(\x,\y), \quad \y \in \de \Om.
 \label{KH}
\end{align}
Here $E(\x, \y)$ is proportional to the fundamental solution of the Helmholtz equation:
\begin{align}
 E(\x, \y) =  \frac{\cos k |\x - \y|}{|\x - \y|} ; \quad  \left( \Delta  + k^2 \right) E  = -4\pi\delta (\x - \y),
 \label{E}
\end{align}
$\n$ is the exterior to $\Om$ unit normal vector and $F$ is the solution of the Dirichlet problem in the ball $B_R$:
\begin{align}
\label{F}
\Delta F + k^2 F &= 0, \quad F \in H^2 (B_R ); \quad
F(\x,\y) = \frac{\de E(\x,\y)}{\de \n_\y}, \quad \x \in \de B_R.
\end{align} 
Then \rf{eq_ut} and the second relation in \rf{bc_ut} hold for the function \rf{ut}. Since the kernel $H(\x,\y)$ differs from the kernel of the standard double layer potential by a continuous function, the limiting value of $\ut$ on $\de \Om$ from outside of $\Om$ is equal to $2\pi  \alpha + \H \alpha$,
and the first condition in \rf{bc_ut} takes the form of a Fredholm integral equation
\begin{align}
 \left(2\pi \Id + \H \right) \alpha = -u_0, \quad \x \in \de \Om.
 \label{fe}
\end{align}
Hence  the problem \rf{eq_ut},\rf{bc_ut}  is equivalent to
\rf{ut},\rf{fe}.

To find the asymptotic solution of \rf{fe} we introduce a new variable $\bxi = \x/a$ in which $\Om = \Om(a)$ becomes an $a$-independent domain $\Omh$. In what follows, functions and operators in new variables are marked by the ``hat''. In new variables \rf{fe} has the form
\begin{align}
 \left(2\pi \Ih + \Hh \right) \walpha = -\what{u}_0, \quad \bxi \in \de \Omh.
 \label{feh}
\end{align}
Here and below operators are denoted by bold capital letters and their integral kernels by lightfaced counterparts.
\begin{lemma}
\label{Hexp}
Operator $\Hh$ is a $C^\infty$ function of $a$ and can be expanded for small $a$ in the Taylor series
\begin{align}
 \Hh &\sim \Hh_0 + \sum_{n=2}^\infty \Hh_n a^n, 
 \label{H_exp}
\end{align}
where $ \Hh_n$ are the Taylor coefficients of $\Hh$, the series \rf{H_exp} and all its derivatives in $a$ converge asymptotically in the operator norm to the corresponding derivatives of $\Hh$, and
$$\what{H}_0 (\bxi, \et) = \frac{(\bxi - \et )\cdot \n_{\et}}{|\bxi - \et|^3},~~\what{H}_2 (\bxi, \et) = \frac{k^2(\bxi - \et)\cdot \n_{\et}}{2\, | \bxi - \et|} , ~~~\et = \y/a.$$
\end{lemma}
\begin{proof} 
We represent $ \Hh$ in the form $ \Hh=\what{\Es}'-\what{\F}$  by splitting the integral in \rf{H} in two terms using \rf{KH}. The explicit formula \rf{E} and the relation $dS_\y=a^2dS_\et$ allow one to expand operator $\what{\Es}'$ in a power series, and this series has the same form as \rf{H_exp} for $\Hh$, with the same main coefficients for $a^0$ and $a^2$. It remains to show that subtraction of $\what{\F}$ does not change the formula \rf{H_exp} for $\Hh$. 

The boundary condition in $\rf{F}$ is infinitely smooth in $\x$ and $\y$ when $|\y|<Ca,~\x\in \de B_R$. Thus $F\in C^\infty$ when $|\y|<Ca,~|\x|\leq R$, and this together with $dS_\y=a^2dS_\et$ imply that $\what{\F}\sim \sum_{j\geq2}\what{\F}_ja^j,~ a\to0.$ 
We still need to demonstrate that $\what{\F}_2=0$, which we can accomplish by proving that $F(\z,\z)=0$.
The latter relation follows from  $\rf{F}$ since the boundary condition there is an odd function of $\x$ when $\y=0$ and therefore, $F(\x,\z)$ is odd.
\end{proof}

In order to study equation \rf{feh} we start with the equation where the operator $\Hh$ is replaced by its main term of asymptotics $\Hh_0$ and $\what{u}_0$ is replaced by an arbitrary function $\what{g}$:
\begin{align} \label{h0}
\left(2\pi \Ih + \Hh_0\right) \walpha &= \what{g}.
\end{align}

Equation \rf{h0} appears if one looks for the solution of the exterior Dirichlet problem for the harmonic  function $V(\bxi)$
\begin{align}
\label{DV}
 \Delta V &= 0, \quad \bxi \in \R^3 \setminus \Omh, \quad V \in H^2_{\rm loc} (\R^3 \setminus \Omh), \\[2mm]
 V &= \O\left(|\bxi |^{-1} \right), \quad |\bxi| \to \infty, \\[2mm]
 \left. V \right|_{\de \Omh} &= \what{g} \in H^\oh (\de \Omh),
 \label{h}
\end{align}
in the form of a double layer potential with the density  $\walpha$, i.e.
\begin{align}
 V = \int_{\de \Omh} \walpha(\et)\,\frac{\de}{\de \n_\et} \frac{1}{|\bxi - \et|}\,  \D S_\et
 =\Hh_0 \,\walpha(\et).
\end{align} 

\begin{lemma}
 Equation \rf{feh} is uniquely solvable for small values of $a>0$ and can be presented
in $H^{ \frac{1}{2}} (\de \Omh)$ as a power series
 \begin{align}\label{aa}
  \walpha(\bxi) \sim 
  a^{-2}\sum_{n=0}^\infty \walpha_n(\bxi)\, a^n, \quad a \to 0, \quad \bxi \in \de \Omh,
 \end{align}
where $\walpha_0, \walpha_1$ are constants and $\walpha_0 =  \dfrac{q u_0 (\z)}{|\Omh|  k^2}$.
 \label{lemma_ser}
\end{lemma}
\begin{proof}
It is well known that the problem \rf{DV}-\rf{h} is uniquely solvable but the solution cannot always be found in the form of a double layer potential. The Fredholm operator in \rf{h0} (in the case of the {\it exterior} Dirichlet problem) has a one-dimensional kernel $\K$ and cokernel $\C$. The kernel consists of constant functions and $\what{u}_1(\bxi)=1, ~\bxi \in \de \Omh,$ will be chosen as a basis there. The cokernel $\C$  consists of the densities (of the charge) for which a single layer potential is constant on $\de\Omega$. Hence, we can choose the basis in the cokernel as $\what{u}_2(\bxi)=\frac{\de}{\de \n_\bxi}\, v, ~\bxi\in \de\Omh,$ where $v$ is the solution of \rf{v} which is, in fact, the same problem as \rf{DV}-\rf{h} with $\what{g} = 1$:

We represent the domain of the operator in \rf{feh} in the vector form $\walpha = (\gamma\what{u}_1, \what{u}_{1\perp})$, where the first component is the $L^2$-projection of $\walpha$ on $\K$, and $\what{u}_{1\perp}$ is orthogonal to $\what{u}_1$ in $L^2$, i.e. $\gamma = \frac{1}{|\de \Omh|} \int_{\de \Omh} \walpha\,(\bxi) \D S_\bxi$. Hereafter all norms are taken in $L^2(\de\Omh)$. Similarly, we represent the functions $\what{g}$ from the range of the operator in the form $\what{g} = (\beta, \beta_\perp)$, where $\beta$ is the $L^2$-projection of $\what{g}$ on $\C$, and $\beta_\perp$ is orthogonal to $\beta$ in $L^2$.
Since $ 2\pi\Ih + \Hh_0$ acts only in orthogonal compliments to $\C$ and $\K$, Lemma \ref{Hexp} and 
the $C^\infty$ dependence in $a$ of the right-hand side $-\what{u}_0 = -u_0( a \bxi)$ in \rf{feh} lead to the following matrix representation of equation \rf{feh}:
\begin{align}
 \left(
 \begin{array}{cc}
  a^2 \Th_2 + a^3 \Th_3 + \O(a^4) & \O(a^2) \\[2mm]
  \O(a^2) &  2\pi\Ih + \Hh_0 +\O(a^2)
 \end{array}
\right)
\left(
\begin{array}{c}
 \gamma\what{u}_1 \\[2mm]
\what{u}_{1\perp}
\end{array}
\right) =
 \left(
\begin{array}{c}
 \left(c_1   + c_2a   +\O(a^2) \right) \what{u}_2 \\[2mm]
 \beta_{0\perp} + \O(a)
\end{array}
\right),
 \label{T}
\end{align}
where the constant $\gamma=\gamma(a)$ and the function $\what{u}_{1\perp}=\what{u}_{1\perp}(a,\bxi)$ are unknown and depend on $a$, and all the remainders can be expanded in a Taylor series in $a$ but for transparency, we provide only the leading terms. Here $\beta_{0\perp}$ is the projection of $-u_0(\z)$ into the space of functions orthogonal to $\what{u}_2$ and $\Th_i \what{u}_1 = \Hh_i \what{u}_1,~i=2,3,$  projected on $\what{u}_2$. Constants $c_i,~i=1,2, $ are the leading coefficients of the Taylor expansion of the projection of the right-hand side in \rf{feh} on $\what{u}_2$. In particular,
\begin{align}
 \Th _2\, \what{u}_1 = d\,  \what{u}_2, \quad    d 
 = \frac{k^2}{2 \|\what{u}_2\|^2} \int_{\de \Omh} \int_{\de \Omh} \frac{(\bxi - \et)\cdot \n_{\et}}{| \bxi - \et|}\, \what{u}_2 (\bxi) \,\D S_\et \, \D S_\bxi,
 \label{Th}
\end{align}
and
\begin{align}
 c_1 &= -\frac{u_0 (\z)}{\|\what{u}_2\|^2} \int_{\de \Omh} \what{u}_2\, \D S_\bxi 
 = -\frac{u_0 (\z)}{\|\what{u}_2\|^2} \int_{\de \Omh} \frac{\de v}{\de \n}\, \D S_\bxi =
 \frac{4\pi q u_0 (\z)}{\|\what{u}_2\|^2}. 
 \label{c}
\end{align}

We will show below that the coefficient $d$ in \rf{Th} is not zero. Thus after multiplication of the first row in \rf{T} by $a^{-2}$, the main term of asymptotics of the matrix operator becomes invertible, and the inverse of the matrix operator can be expanded in a power series in $a$. This leads to  \rf{aa} with the coefficient for $a^{-2}$ equals $\frac{c_1}{d}$. It remains only to show that $\frac{c_1}{d}=\frac{q u_0 (\z)}{|\Omh|  k^2}.$

Let us evaluate $d$. For the inner integral in \rf{Th}, we have
\begin{align*}
  \int_{\de \Omh} \frac{(\bxi - \et)\cdot \n_{\et}}{| \bxi - \et|}\, \D S_\et =
  \int_{\Omh} \nabla_\et \cdot \frac{\bxi - \et}{| \bxi - \et|}\, \D \et
  = -2 \int_{\Omh} \frac{ \D \et}{| \bxi - \et|} = w(\bxi),
\end{align*}
where  $\Delta w = 8\pi $
in $\Omh$,  $\Delta w = 0$ in $\R^3 \setminus \Omh$, $w(\bxi)=\O(|\bxi|^{-1})$ at infinity. Therefore
\begin{align*}
  d  = \frac{k^2}{2 \|\what{u}_2\|^2} \int_{\de \Omh} w(\bxi)\, \what{u}_2 (\bxi)  \, \D S_\bxi
 =  \frac{k^2}{2 \|\what{u}_2\|^2} \int_{\de \Omh}w(\bxi)\,\frac{\de v(\bxi)}{\de \n} \, \D S_\bxi 
 = \frac{k^2}{2 \|\what{u}_2\|^2} \int_{\de \Omh}\frac{\de w(\bxi)}{\de \n} \,v(\bxi)\, \D S_\bxi,
\end{align*}
where $v$ is defined in \rf{v} and the last equality is obtained from the Green formula in $\R^3 \setminus \Omh$ in which integration over the large sphere does not contribute to the formula due to a fast enough decay of functions $v,w$ at infinity. Now we can omit $v$ from the last integrand since $v=1$ on $\de \Omh.$ Then the integral is equal to $4\pi | \Omh|$, and therefore
\begin{align*}
 d
 =  \frac{4\pi |\Omh|k^2}{\|\what{u}_2\|^2}.
\end{align*}
This and \rf{c} imply the validity of the first coefficient in the right-hand side of \rf{aa}.

\end{proof}

The following statement is a direct consequence of Lemma \ref{lemma_ser}  and the estimate
\[
\|\alpha\|_{L_1 (\de\Om)}=a^2\|\walpha\|_{L_1(\de\Omh)}\leq Ca^2\|\walpha\|_{L_2(\de\Omh)}\leq Ca^2\|\walpha\|_{H^{ \frac{1}{2}}(\de\Omh)}.
\]
\begin{lemma}
 Equation \rf{fe} is uniquely solvable in $H^{ \frac{1}{2}} (\de \Om)$ for small values of $a>0$ and its solution can be presented in $L_1(\de\Om)$
as a power series
 \begin{align}\label{aa1}
  \alpha = \alpha(a,\x) \sim \sum_{n=0}^\infty \alpha_n \left(\x/a\right)\, a^{n-2}, \quad a \to 0, \quad \alpha_n \left(\x/a\right) = \walpha_n (\bxi) , 
 \end{align}
 where $\alpha_0=\walpha_0,  ~ \alpha_1=\walpha_1$ are constants and dependence of $\alpha$ on $a$ is explicitly included in the argument.
 \label{lemma_ser1}
\end{lemma}
\noin
{\bf Proof of Theorem \ref{t1}.}
Function  $\ut$ has the form \rf{ut}, where $\alpha$ satisfies \rf{fe}, and the solution of the latter equation is described in Lemma \ref{lemma_ser1}. We split $\ut$ in two terms
\begin{align}
\ut=\int_{\de\Om}\frac{\de E(\x,\y)}{\de \n_\y}\,\alpha(a,\y)\,\D S_\y-\int_{\de\Om}F(\x,\y)\alpha(a,\y)\,\D S_\y:=v_1-v_2
\label{v1}
\end{align}
which correspond to the representation of $H$ in \rf{KH} as a difference of two terms.

Let us evaluate $v_1$. Formula  \rf{E}  implies 
\[
\frac{\de E(\x,\y)}{\de \n_\y}=\left(\frac{k\sin k |\x-\y|}{|\x-\y|^2}+\frac{\cos k|\x-\y|}{|\x-\y|^3}\right)(\x-\y) \cdot \n_\y.
\]
We expand the right-hand side in a power series in $\y$ when $|\y|<Ca\ll 1, ~|\x| \geq\delta$ (preserving $\n_\y$) and obtain
\[
\frac{\de E(\x,\y)}{\de \n_\y}=\sum_{j=0}^\infty \Eb_j(\x,\y) \cdot \n_\y,
\]
where $\Eb_j$ are vectors whose components are homogeneous polynomials in $\y$ of order $j$ with infinitely smooth in $\x$ coefficients, and 
\[
\Eb_0 \cdot \n_\y=\left(\frac{k\sin k|\x|}{|\x|^2}+\frac{\cos k|\x|}{|\x|^3}\right) \x \cdot\n_\y,
\]
\[
\Eb_1 \cdot \n_\y=\left(-\frac{k^2\cos k|\x|}{|\x|^3}+3\frac{k\sin k|\x|}{|\x|^4}+3\frac{\cos k|\x|}{|\x|^5}\right)(\x\cdot\y)(\x \cdot\n_\y)
\]
\[
-\left(\frac{k\sin k|\x|}{|\x|^2}+\frac{\cos k|\x|}{|\x|^3}\right)\y \cdot \n_\y:=
e_{11}-e_{12}.
\]
Since $|\de \Om| = c a^2$, where $c$ is a constant, from \rf{v1} and Lemma \ref{lemma_ser1} it follows that
\begin{equation}
v_1=\sum_{j=0}^\infty v_{1j}(\x)\,a^j, \quad a\to 0,  \quad \delta\leq|\x|\leq R,
\label{v1_ser}
\end{equation}
where the coefficients are infinitely smooth in $\x$ and series converges in $C^\infty$. The aforementioned formula is apparent if one changes the variables $\y \to a \et$ in \rf{v1}.

Let us evaluate the first two coefficients in the series above. 
The divergence theorem implies that
\begin{align*}
v_{10}&=\alpha_0\, a^{-2}\int_{\de\Om}\Eb_0 \cdot \n_\y \,\D S_\y=0,\\[2mm]
\int_{\de\Om}e_{11}\,\D S_\y&=\left(-\frac{k^2\cos k|\x|}{|\x|}+3\frac{k\sin k|\x|}{|\x|^2}+3\frac{\cos k|\x|}{|\x|^3}\right)|\Om|,\\[2mm]
\int_{\de\Om}e_{12}\,\D S_\y&=\left(3\frac{k\sin k|\x|}{|\x|^2}+3\frac{\cos k|\x|}{|\x|^3}\right)|\Om|.
\end{align*}
Therefore the second term $v_{11} a$ in \rf{v1_ser} equals
\begin{align}
v_{11}a &=\alpha_1\, a^{-1}\int_{\de\Om}\Eb_0 \cdot\n_\y\,\D S_\y
+\alpha_0\, a^{-2} \int_{\de\Om}\Eb_1 \cdot \n_\y\,\D S_\y = \alpha_0\, a^{-2} \int_{\de\Om}(e_{11} - e_{12})\,\D S_\y \nonu \\[2mm]
&= -\alpha_0 \, a^{-2} |\Om| \,\frac{k^2\cos k|\x|}{|\x|}
=-\walpha_0\, a |\Omh| \,\frac{k^2\cos k|\x|}{|\x|}.
\end{align}
Hence,
\begin{equation}\label{v11}
v_1=\sum_{j=1}^\infty v_{1j}(\x)\,a^j, \quad a\to 0,  \quad \delta\leq|\x|\leq R, \quad
v_{11}=-\walpha_0 |\Omh| \,\frac{k^2\cos k|\x|}{|\x|}.
\end{equation}

From \rf{F} it follows that $v_2$ is the solution of the following problem in the ball $B_R$
\begin{align}
\Delta v_2 + k^2v_2 &= 0, \quad v_2 \in H^2 (B_R ); \quad
v_2 =v_1, \quad \x \in \de B_R.
\end{align}
Thus,
\begin{align}
v_2=\sum_{j=1}^\infty v_{2j}(\x)\,a^j, \quad a\to 0,  \quad |\x|\leq R.
\label{v2}
\end{align}
Function $v_{21}$ is a smooth solution of the Helmholtz equation in the ball 
and $v_{21}=-\walpha_0|\Omh| \,\dfrac{k^2\cos kR}{R}$ on the boundary, and therefore 
\begin{align}
 v_{21} = -\walpha_0 k^2 |\Omh| \cot kR\, \frac{\sin k |\x|}{|\x|}.
 \label{v21}
\end{align}

Since $\ut=v_1-v_2$ formulas \rf{v11}, \rf{v2}, \rf{v21} imply formula for $\ut$ in the
statement of the theorem from which the formula for the derivative immediately follows.
\qed

\section{Asymptotics of the interior DtN map in the basis $\psi_{\mathscr E}, \psi_{\perp}$}
\label{main}
\setcounter{equation}{0}

Recall that Lemma \ref{l3a} provides the basis $\psi_s, 1 \leq s \leq n,$ of the finite-dimensional kernel $\mathscr E$ of the exterior operator 
\[\Np_{\bk,0}-\Nm_{0,0}: \Ht(\de B_R) \to\Ho (\de B_R).\]
This operator has a simple matrix representation \rf{A} if the domain of the operator and its range are presented as a direct sum of the $L^2$-projections $\psi_{\mathscr E}$ on $\mathscr E$ and function $\psi_{\bot}$ which is orthogonal in $L^2(\de B_R)$ to $\mathscr E$. The formula \rf{conc} provides a matrix representation for the perturbed exterior DtN operator. We need a similar matrix representation for the interior operator.
We need to specify only the upper left element of the latter matrix, i.e. we need the asymptotic behavior of 
\begin{align}
 \left( \left(\Nm_{a,\ep}-\Nm_{0,\ep}\right) \psi_{i}, \psi_{j} \right), \quad \hpsi_{i}, \hpsi_{j} \in \mathscr E,
 \label{M}
\end{align}
as $\ep, a \to 0$. The matrix of this form in the basis $\hpsi_s, 1 \leq s \leq n,$ will be denoted by $\M=\M(a,\ep)$.
We will use the same notation for the operator in the space $\mathscr E$ defined by the matrix $\M$.

\begin{theorem}
\label{t2}
If $R \notin \{R_i\}$ then the matrix $\M$ have the form
\begin{align}
\M=\M(a,\varepsilon) =\M(a,0)+\O(a|\varepsilon|), \quad a, \varepsilon \to 0,\quad  \M(a,0)=4\pi a q\, \J + \O (a^2 ), 
 \label{Mat2}
\end{align}
where $\J$ is an $n\! \times \! n$ all-ones matrix.
\end{theorem}
\begin{proof}
Since the power expansion of function $\ut$ in \rf{uwa} starts with the first power of $a$, the first relation in \rf{Mat2} follows from the first equality in \rf{Nam}. 

In order to obtain the second relation in \rf{Mat2}, we note that the solution $u_0$ of \rf{u0} with $\psi = \psi_i$ equals $u_0 = \E^{-\I (\bk - \mb_i) \cdot \x} + \O(|\ep|)$  and therefore $u_0 (\z) = 1 + \O(|\ep|)$. Thus, \rf{Nam} leads to
\begin{align}
\left(\Nm_{a,\varepsilon}-\Nm_{0,\varepsilon}\right) \psi_i |_{\ep=0}=\frac{a q k }{R\sin (kR)} + \O (a^2 ), \quad a \to 0, \quad \x\in \de B_R.
 \label{dif}
\end{align}
Using the value of the integral
\[
\int_{\de B_R} \E^{\I (\bk-\mb_j) \cdot \x}\, \D S=4\pi R^2\, \frac{\sin (|\bk-\mb_j| R)}{|\bk-\mb_j| R} =
4\pi R\, \frac{\sin (|\bk| R)}{|\bk|} 
= \frac{4\pi R}{k}\,\sin(kR) + \O(|\ep|),
\]
and \rf{dif} we obtain for the entries of matrix $\M$
\[
 M_{i,j}(a,0) = \left(\left(\Nm_{a,\varepsilon}-\Nm_{0,\varepsilon}\right) \psi_i, \psi_j \right) |_{\ep=0} =\frac{a q k}{R\sin (kR)} \frac{4\pi R}{k}\,\sin(kR) + \O (a^2)
 = 4\pi a q + \O (a^2 ).
\]
\end{proof}

\section{Derivation of the dispersion relations and clusters of Bloch waves}
\label{pt1}
\setcounter{equation}{0}

We recall that in the Dirichlet problem propagation frequency, $\om$ starts from a nonzero value $\om_c$ called the cutoff frequency. The wavelength corresponding to $\om_c$ is denoted by $\lambda_{\max} = 2\pi c/\om_c$.
\begin{theorem}
 \label{t3}
 If $\bk$ is a non-exceptional Bloch vector then solution of the problem \rf{Hz1},\rf{bc1a} has the form
 \begin{align}
 \label{nes}
 u(\x) = C  \left(\E^{-\I \bk \cdot \x}  + \ut(\x)\right), \quad
 \rrbracket  \E^{\I \bk \cdot \x}\, \ut \llbracket = 0, \quad \| \ut \|_{C^m (\Pi \setminus B_R)} = \O(a), 
 \quad a \to 0,
\end{align}
where $C$ is a constant and $m$ is arbitrary. The wave number $k$ is separated from zero. The dispersion relation 
and the maximum wavelength $\lambda_{\max} = \dfrac{2\pi}{k_{\min}}$ (the cutoff wavelength) supported by the periodic medium with cavities are given by
 \begin{align}
  \label{dr1}
 k^2  &= |\bk|^2  + \frac{4\pi a q}{|\Pi|} + \O(a^2), \quad a \to 0, \\[2mm]
 \lambda_{\max} &= \sqrt{\frac{\pi |\Pi|}{aq}}\left(1 + \O (a) \right), \quad a \to 0.
 \label{lam_max1}
\end{align}
The coefficient $q$ depends on the shape of the cavity and is determined by \rf{v}.
\end{theorem}
\begin{remark}
 Solution $u(\x)$ has the form of a perturbed plane wave propagating in the direction of vector $\bk$. Due to the Bloch condition in \rf{nes}, the perturbation $\ut$ has order $O(a)$ uniformly in $\x \in \R^3$ outside of a neighborhood of cavities.
\end{remark}

\begin{theorem}
 \label{t4}
 Let $\bk$ be an exceptional Bloch vector of order $n \geq 2$, $\mb_j\in\mathbb Z^3_b, 1 \leq j\leq n, |\bk-\mb_j|=|\bk|$. 
Then for small $a>0$, there are  $n'\geq 1$ distinct values of $k=k_s,~1\leq s\leq n',$ such that $k_s\to|\bk|$ as $a\to0$ and problem \rf{Hz1},\rf{bc1a} with $k=k_s$ has a non-trivial solution. These solutions have the form of clusters $u_s$ of perturbed plane waves propagating in directions $\bk-\mb_j,~1 \leq j\leq n$. 
 
One of them has the form
 \begin{align}\label{1u}
 u_1(\x) = C_1 \left( \sum_{j=1}^n \E^{-\I (\bk - \mb_j) \cdot \x} + \ut_{1}\right), \quad \rrbracket  \E^{\I \bk \cdot \x}\, \ut_{1} \llbracket = 0, \quad \| \ut_{1} \|_{C^m (\Pi \setminus B_R)} = \O(a),
 \quad a \to 0,
\end{align}
where $C_1$ is a constant  and $m$ is arbitrary. Its dispersion relation is given by
 \begin{align}\label{1uk}
  k_1^2 = |\bk|^2 + \frac{4\pi a q n}{|\Pi|} + \O(a^2), \quad a \to 0
 \end{align}
 with the cutoff wavelength
 \begin{align}
  \lambda_{\max} &= \sqrt{\frac{\pi |\Pi|}{aqn}}\left(1 + \O (a) \right), \quad a \to 0.
 \end{align}
If $n=2$, then
\begin{align}\label{2u}
 u_2(\x) = C_2 \left( \E^{-\I (\bk - \mb_1) \cdot \x} - \E^{-\I (\bk - \mb_2) \cdot \x} + \ut_{2}\right), \quad \rrbracket  \E^{\I \bk \cdot \x}\, \ut_{2} \llbracket = 0, \quad \| \ut_{2} \|_{C^m (\Pi \setminus B_R)} = \O(a),
 \quad a \to 0,
\end{align}
where $C_2$ is a constant  and $m$ is arbitrary.

The terms of order $a$ in the dispersion relations of all the clusters $u_s$ with $s>1$ are vanishing, and these dispersion relations are
\begin{align}\label{2uk}
  k_s^2 = |\bk|^2 + \O(a^2), \quad a \to 0, \quad s>1,
 \end{align} 
with the cutoff wavelength $\lambda_{\max} = \O \left(a^{-1}\right)$.

Clusters $u_s$ cannot consist of a wave propagating in one specific direction $\bk-\mb_j,~1 \leq j\leq n$.
\end{theorem}
\begin{remark}
One could try to construct a linear combination of solutions of \rf{Hz1},\rf{bc1a} with different values of $k=k_s$ to obtain a wave propagating in one direction. However, such a construction does not have mathematical or physical meaning since the terms would satisfy different equations with the different time frequencies $\om_s = k_s/c, ~s\geq 1$. 
\end{remark}
\begin{remark} The arguments below allow one to obtain the complete Taylor series in $a$ for the Bloch waves \rf{nes}, \rf{dr1} in Theorem \ref{t3} and for the first cluster \rf{1u}, \rf{1uk} in Theorem \ref{t4}. 
\end{remark}
We will prove only Theorem \ref{t4} since its proof is applicable to validate Theorem \ref{t3} simply by putting $n=1$. 
\begin{proof}
We can express equation \rf{nn} as
\begin{equation}
\label{matr}
 (\Np_{\bk,\varepsilon}-\Nm_{0,\varepsilon})\psi - (\Nm_{a,\varepsilon}-\Nm_{0,\varepsilon})\psi=0.
\end{equation}
Operators in \rf{matr} act from the space  $\Ht(\de B_R)$ to $\Ho (\de B_R)$. According to Lemma \ref{one-to-one}, we can determine the dispersion relation by finding the values of the parameters $\ep$ and $a$ that allow equation \rf{matr} to have a non-trivial solution.

To achieve this, we divide the domain and range of operators into two orthogonal components in $L^2(\de B_R)$. The first component, $\mathscr E$, is a finite-dimensional space of functions spanned by functions $\psi_{s}$ described in Lemma \ref{l3a}. By doing this, we can rewrite \rf{matr} in a matrix form similar to that in \rf{A} and \rf{conc}. Specifically, for any $\bk$, let $\psi = (\psi_{\mathscr E}, \psi_{\bot})$ be the vector representation of function $\psi \in \Ht (\de B_R)$, where $\psi_{\mathscr E}$ is the $L^2$-projection of $\psi$ into $\mathscr E$. If the point $\bk$ is not exceptional, then $\mathscr E$ is a one-dimensional space of functions proportional to $\E^{-\I \bk \cdot \x}$. Based on \rf{conc} and the expression \rf{k-egv} for $\ep$, we can express equation \rf{matr} as follows:

\begin{equation}\label{matr1}
  \left(
 \begin{array}{cc}
  \Csbi\ei + \ei^2 \Dsbi_{11}-\B_{11}(a,\varepsilon) & \ei \Dsbi_{12}-\B_{12}(a,\varepsilon) \\[2mm]
  \ei \Dsbi_{21}-\B_{21}(a,\varepsilon) & \A + \ei \Dsbi_{22}-\B_{22}(a,\varepsilon)
 \end{array}
 \right)
  \left(
 \begin{array}{cc}
  \psi_{\mathscr E} \\[2mm]
  \psi_{\bot}
 \end{array}
 \right)=0, \quad k = (1 +  \ep) |\bk|,
\end{equation}
where matrix elements $\B_{i,j}$ are defined by the operator $\Nm_{a,\varepsilon}-\Nm_{0,\varepsilon}$. Theorem \ref{t2} implies that $\|\B_{ij}\|=\O(a),~ a,\varepsilon\to 0.$ The element $\B_{1,1}$ coincides with the operator $\M=\M(a,\varepsilon)$ whose entries are given by \rf{M}.

We will solve the system of equations in \rf{matr1} for $\psi_{\mathscr E}$ and $\psi_{\bot}$ and find $\ei = \ei (a)$ for which this system has a non-trivial solution. Then the equation for $k$ in \rf{matr1} gives the relation $\kp = \kp (\ei)$ for which 
\rf{matr},\rf{matr1} and therefore \rf{Hz1},\rf{bc1a} have non-trivial solutions. Thus $k = k(\ep( a))$ is the dispersion relation.

Since operator $\A$ is invertible, the second equation of the system \rf{matr1}  for $\psi_{\mathscr E}$ and $\psi_{\bot}$ can be solved for $\psi_{\bot}$ yielding
\begin{align}
 \label{phi2_norm}
 \|\psi_{\bot}\| \leq c (a+|\ei|) \|\psi_{\mathscr E} \|.
\end{align}
This reduces \rf{matr},\rf{matr1} to an equation in the finite-dimensional space $\mathscr E$:
\begin{align*}
\label{kpe}
 [\Csbi\ei-\Msbi( a,\varepsilon)+O(a^2+\ei^2)]\psi_{\mathscr E}=0.
\end{align*}
We substitute here formulas \rf{C} and \rf{Mat2} for $\Csbi$
and $\Msbi$, and obtain
\begin{equation}\label{drel}
\left[ 2\ei\Im-\frac{4\pi a q}{|\bk|^2 |\Pi|}\, \J+\O\left(a^2 + \ei^2\right)\right]\psi_{\mathscr E}=0.
\end{equation}
Since $\psi_{\mathscr E} = \sum_{s=1}^n \tau_s \psi_s$, formula \rf{drel} can be rewritten in terms of ${\bm \tau} = (\tau_1, \ldots, \tau_n)$:
\begin{equation}\label{drel1}
\left[ 2\ei\Im-\frac{4\pi a q}{|\bk|^2 |\Pi|}\, \J +\O\left(a^2 + \ei^2\right)\right]{\bm \tau}=0.
\end{equation}

Equation \rf{drel1} cannot have non-trivial solutions ${\bm \tau}$ for small $a,\ei,$  if $|\ei| \gg a$. Thus, $|\ep| \leq Ca$ when $a\ll 1$. We substitute $\ep$ by $\ep = \dfrac{2\pi a q \hat{\ep}}{ |\bk|^2 |\Pi|}$ in \rf{drel1} and divide the equation by $a$. This leads to
\begin{equation}\label{drel2}
\left[ \hat{\ep}\Im- \J +\O\left(a + a\hat{\ep}^2\right)\right]{\bm \tau}=0, \quad a\ll 1, \quad |\hat{\ep}|\leq C.
\end{equation}

Matrix $\J$ has a simple eigenvalue $\nu_1 = n$ with the eigenvector ${\bm \mu}_1=(1,\ldots,1)$, and its other eigenvalues are zeros: $\nu_s = 0$, $2 \leq s \leq n$. Equation \rf{drel2} cannot have non-trivial solutions ${\bm \tau}$ unless $|\hat{\ep} -n| = \O(a)$ or $|\hat{\ep}| = \O(a)$. The implicit function theorem implies that \rf{drel2} has a unique infinitely smooth in $a$ solution $\hat{\ep}=\hat{\ep}_1 (a), ~{\bm \tau}={\bm \tau}_1(a) $ with $\hat{\ep}_1 (0)=1, ~{\bm \tau}_1(0)= (1,\ldots,1)$. For all other bounded for small $a$ solutions $\hat{\ep} $ of \rf{drel2} we have $|\hat{\ep}|=O(a).$ Hence, one (infinitely smooth in $a$) solution of \rf{drel1} has the form
\begin{align}\label{2ei}
2\ep_1=\frac{4\pi a qn}{|\bk|^2 |\Pi|} +\O(a^2), \quad {\bm \tau_1}=C_1({\bm \mu}_1 +\O(a)), \quad a\to 0, 
\end{align}
and 
\begin{align}\label{2ei5}
\ep_s=\O(a^2), ~s> 1,
\end{align}
for all other solutions of \rf{drel1}. If $n=2$ then zero eigenvalue of $\J$ is also simple, and therefore \rf{drel2} has another infinitely smooth in $a$ solution $\hat{\ep}=\hat{\ep}_2 (a), ~{\bm \tau}={\bm \tau}_2(a) $ such that $\hat{\ep}_2 (0)=0, ~{\bm \tau}_2(0)= (1,-1)$. 

Next, the relation for $k$ in \rf{matr1} implies two equalities: $\dst 2 = (k + |\bk|)/|\bk|  - \ep$ and $\ep = (k - |\bk|)/|\bk|$, from which we obtain
\begin{align}
 2\ei =\frac{(k -|\bk|)(k +|\bk|)}{|\bk|^2} + \O(\ep^2) = \frac{k^2 - |\bk|^2}{|\bk|^2}  + \O(\ep^2) = \frac{k^2 - |\bk|^2}{|\bk|^2}  + \O(a^2),
 \label{Ce}
\end{align}
since $|\ep| \leq Ca$. Relations \rf{2ei}-\rf{Ce} will justify \rf{1u},\rf{1uk} and \rf{2u},\rf{2uk} if we prove the estimate for $\ut_1$ in \rf{1u} (the estimate for $\ut_2$ can be obtained similarly).

Recall that function $u_1$ in the domain $\Pi \setminus B_R$ satisfies
\begin{align}
 \label{vv}
 \left( \Delta + k^2 \right) u_1 &= 0, \quad  \x\in \Pi \smallsetminus B_R,~~  \left\rrbracket  \E^{\I \bk \cdot \x} u_1(\x) \right\llbracket =0, ~~ \left. u_1\right|_{r=R}=\psi,
 \end{align}
where $\psi$  is the solution of \rf{matr} and \rf{matr1} for which $\psi_\bot=O(a)$ and 
\begin{align*}
\psi_{\mathscr E}= \sum_{j=1}^n \tau_{1,j}\, \E^{-\I (\bk - \mb_j) \cdot \x} ,
\end{align*}
where $\tau_{1,j}$ are components of the vector ${\bm \tau}_1$ defined in \rf{2ei}. Since \rf{1u} implies that
$$\ut_1 = C_1^{-1} u_1 - \sum_{j=1}^n \E^{-\I (\bk - \mb_j) \cdot \x},$$  
$\ut_1$ is the solution to the problem
\begin{align}
 \label{vv1}
 \left( \Delta + k^2 \right) \ut_1 &= C_1 (|\bk|^2 - k^2) \sum_{j=1}^n \E^{-\I (\bk - \mb_j) \cdot \x}, \quad  \x\in \Pi \smallsetminus B_R,~~  \left\rrbracket  \E^{\I \bk \cdot \x} \ut_1(\x) \right\llbracket =0, ~~ \left. \ut_1\right|_{r=R}=\tilde{\phi},
 \end{align}
where $\tilde{\phi} = \O(a) \cdot \sum_{j=1}^n \E^{-\I (\bk - \mb_j) \cdot \x} + C_1^{-1} \psi_\perp$. From the latter relation and \rf{phi2_norm} where the norm in the left-hand side is taken in the space $\Ht(\de B_R)$, we obtain
 
\begin{align*}
\| \tilde{\phi}\|_{\Ht(\de B_R)} = \O(a), \quad a \to 0.
\end{align*}

Denote the right-hand side in the equation \rf{vv1} by $f$. Since $k^2 -|\bk|^2=O(a)$ (see \rf{2ei} and the first two relations in \rf{Ce}) we have 
\begin{align*}
 \| f\|_{L^2(\Pi \setminus B_R)} = \O(a).
\end{align*}

Denote by $\U_\ep:L^2(\Pi \setminus B_R) \times \Ht(\de B_R)\to H^{2}(\Pi \setminus B_R)$ the operator that maps arbitrary right-hand sides $(f, \tilde{\phi})$ of problem \rf{vv1} into the solution, and denote by $\U_0$ a similar operator when $ k^2$ in \rf{vv1} is replaced by $|\bk|^2$. Due to the choice of $R\notin \{R_i\}$, the standard a priori estimates for elliptic problems imply that operator $\U_0$ is bounded. Then operator $\U_\ei$ is also bounded when $a\ll 1$ since $k^2 -|\bk|^2=O(a)$. Since the norms of $f$ and $\tilde{\phi}$ have order $\O(a)$, the boundness of operator $\U_\ep$ implies 
\begin{align}
\label{hu2}
\| \ut_{1} \|_{H^2 (\Pi \setminus B_R)} = \O(a), \quad a \to 0.
\end{align}
We could enlarge the domain $\Pi \setminus B_R$ by taking a smaller value $R^\prime \notin \{R_i\}$ instead of $R$, and the estimate \rf{hu2} remains valid in the larger domain.
Then the standard a priori estimates for elliptic equations in subdomains imply that the estimate \rf{hu2} is valid in $C^\infty (\Pi \setminus B_R)$, i.e. the estimate for $\ut_1$ in \rf{1u} holds.This competes the proof of \rf{1u},\rf{1uk},\rf{2u},\rf{2uk}. Formulas for $\lambda_{\max}$ in the statement of Theorem \ref{t4} follow immediately from \rf{1uk},\rf{2uk} if we put there $|\bk| = 0$. Since all the arguments above are valid when $n=1$,
 Theorem \ref{t3} is also proven. It remains only to justify the last statement in Theorem \ref{t4} saying that no cluster may consist of a wave propagating in a single direction when $n>1$.

 If we assume that there is a cluster $u$ propagating in a single direction then there are $\ep$ and ${\bm \tau}$ for which \rf{drel1} is valid and ${\bm \tau}$ has only one nonzero component. One can assume that this component is equal to one. Matrix $\J$ is all-unit and therefore $\J {\bm \tau}$ is not a zero vector. Next, $u \neq u_1$ since cluster $u_1$ contains the waves propagating in all the directions defined by $\E^{-\I (\bk - \mb_j) \cdot \x}, ~1\leq j\leq n$. For all other clusters, the term with $\J {\bm \tau}$ in \rf{drel1} is proportional to $a$ and all other terms have order $\O(a^2)$ due to \rf{2ei5}. Hence, \rf{drel1} cannot be valid.    

\end{proof}

\section{Conclusions}
\label{concl}

We derived and rigorously justify asymptotic expansions for Bloch waves in periodic media containing small cavities of arbitrary shape with Dirichlet boundary conditions. We determined the leading terms of the dispersion relations of Bloch waves and the cutoff frequencies. 
The rigorous approach reveals
that there are exceptional wave vectors $\bk$ for which Bloch waves are a set of clusters of perturbed plane waves propagating in different directions.
We show that for those exceptional wave vectors, no Bloch waves are propagating in one particular direction.
To our knowledge, this effect was mentioned in the literature only in our previous paper and here.

Unlike the transmission, problem \cite{GV:22}, where the second leading term of all the Bloch waves and the dispersion relations are proportional to the concentration of the inclusions, in the case of cavities there are solutions and dispersions with different orders of corrections. These corrections are proportional to the size of the cavity only for one cluster for each exceptional wave vector.
 
There are many important theoretical and applied issues yet to be explored. They include the behavior of the clusters and the dispersion relations in a neighborhood of exceptional Bloch vectors, asymptotic determination of spectral gaps near exceptional points, experimental and numerical observation of cluster waves for the finite size of inclusions, a study of clusters of electromagnetic and elastic waves, and more.
 
\section*{Acknowledgment}

The work of B. Vainberg was supported by the Simons Foundation grant no. 527180.
%
%

%

 \newcommand{\noop}[1]{}

\end{document}